\documentclass[conference]{IEEEtran}
\IEEEoverridecommandlockouts
\usepackage{cite}
\usepackage{amsmath,amssymb,amsfonts}
\usepackage{algorithmic}
\usepackage{graphicx}
\usepackage{textcomp}
\usepackage{xcolor}
\def\BibTeX{{\rm B\kern-.05em{\sc i\kern-.025em b}\kern-.08em
    T\kern-.1667em\lower.7ex\hbox{E}\kern-.125emX}}
\begin{document}

\title{Towards Practical Real-Time Low-Latency Music Source Separation
\thanks{This work was supported by the National Natural Science Foundation of China (Grant No. 62402211) and the Natural Science Foundation of Jiangsu Province (Grant No. BK20241248).}
}
\author{
\IEEEauthorblockN{Junyu Wu,
Jie Liu\thanks{* Jie Liu is the corresponding author
(liujie@nju.edu.cn).}\IEEEauthorrefmark{1},
Tianrui Pan,
Jie Tang,
Gangshan Wu}
\IEEEauthorblockA{State Key Laboratory for Novel Software Technology\\
Nanjing University\\
Nanjing, China \\
221870052@smail.nju.edu.cn,
liujie@nju.edu.cn,
652022330018@smail.nju.edu.cn,
\{tangjie,gswu\}@nju.edu.cn\\
}}

\maketitle

\begin{abstract}
In recent years, significant progress has been made in the field of deep learning for music demixing. However, there has been limited attention on real-time, low-latency music demixing, which holds potential for various applications, such as hearing aids, audio stream remixing, and live performances. Additionally, a notable tendency has emerged towards the development of larger models, limiting their applicability in certain scenarios. In this paper, we introduce a lightweight real-time low-latency model called Real-Time Single-Path TFC-TDF UNET (RT-STT), which is based on the Dual-Path TFC-TDF UNET (DTTNet). In RT-STT, we propose a feature fusion technique based on channel expansion. We also demonstrate the superiority of single-path modeling over dual-path modeling in real-time models. Moreover, we investigate the method of quantization to further reduce inference time. RT-STT exhibits superior performance with significantly fewer parameters and shorter inference times compared to state-of-the-art models. 
\end{abstract}


\section{Introduction}
\label{sec:intro}

Music source separation (MSS) \cite{b1,b2} is a music signal processing technique that aims to isolate different sources from a mixed music signal. The objective sources are typically vocals, drums, bass and other (all the other instruments). This technique is widely used in music remixing \cite{b3,b32}, music transcription \cite{b4}, music information retrieval (MIR) \cite{b5,b33}, and other music-inspired derivative works \cite{b6,b30,b31}.

In recent years, deep learning has significantly advanced the field of MSS. However, most studies \cite{b7,b8,b13,b18} have focused on offline music source separation, using relatively large models, deep networks and non-causal implementations. The high computational cost and large latency make it impossible to apply them to real-time, low-latency scenarios. Real-time, low-latency MSS considers two key factors: algorithmic latency and computational efficiency. Algorithmic latency refers to the required length of context to generate one output frame, while computational efficiency is defined as the processing time per audio frame. Compared to offline MSS, real-time MSS offers immediate feedback and application, thereby enhancing interactivity and user experience.

In this paper, we present a lightweight real-time low-latency model, designated RT-STT, which is based on DTTNet \cite{b13}, an offline MSS model. The contributions are as follows.

Our first contribution is the introduction of feature fusion based on channel expansion. This method can effectively enhance the performance while reducing inference time. We further discuss the reasons behind its effectiveness.

In addition, we demonstrate that single-path modeling is more efficient than dual-path modeling in real-time scenarios, with reduced inference time and slightly improved performance.

Furthermore, we validate the impact of quantization on our model. Despite limited literature on the effects of quantization in audio separation, our study provides useful insights, such as quantization not compromising performance but enhancing inference speed.

Finally, we present RT-STT, achieving a source-to-distortion ratio (SDR) of 5.17 dB on the MUSDB18-HQ dataset \cite{b12}, outperforming the state-of-the-art HS-TasNet (4.65 dB in SDR) with significantly fewer parameters (less than 1 M). Moreover, the inference speed of RT-STT is three times faster than that of HS-TasNet after quantization.  

\section{Related Work}
In areas outside of MSS, there have been more explorations into real-time applications. For example, in speech separation, end-to-end deep learning directly on raw audio has achieved noteworthy outcomes with latency below 5 ms \cite{b14,b11}. Recent advances in low-latency continuous speech separation (CSS) involve splitting long sequential inputs into smaller chunks and iteratively applying intra- and inter-chunk operations, which enable sublinear processing \cite{b34,b35}. Nevertheless, the implementation of low-latency MSS is more complex due to higher sampling rates, the strong dependence on future context, and the potential for deeper networks.
 
Prior to the advent of deep learning, \cite{b25} proposes the Azimuth Discrimination and Resynthesis technique for real-time MSS. 
While not specifically targeting real-time MSS, works such as \cite{b36} have presented architectures that are conducive to real-time applications.
In \cite{b27}, a multi-layer perceptron (MLP) with an algorithmic latency of 23 ms performs singing voice separation. In \cite{b26}, an RPCA-based real-time speech and music separation method is introduced. In \cite{b28}, MMDenseNet is revised to enhance real-time music accompaniment separation, but the latency of approximately 1 second is still too large, making it unsuitable for real-time low-latency application. The latest work (as of this writing) on real-time MSS using deep learning is \cite{b9}. This paper examines some classical models \cite{b10,b11,b14}, and adapts the spectrogram-based X-UMX \cite{b10} and the waveform-based TasNet \cite{b11} for real-time MSS. Subsequently, \cite{b9} integrates the spectral domain into the time-domain Tasnet \cite{b11}, generating a new network called  Hybrid Spectrogram Time-domain Audio Separation Network (HS-TasNet) \cite{b9} which achieves state-of-the-art performance on MUSDB18-HQ dataset \cite{b12}. However, there exist quite a few limitations to their work. Firstly, they consider the adaptation of only two out-of-date models and ignore the recent advances in MSS. For instance, HS-TasNet \cite{b9} employs only the most basic structural and modular components, while the state-of-the-art models \cite{b7,b13,b18} tend to introduce some effective and well-designed modules such as Time-Frequency Convolutions Time-Distributed (TFC-TDF) blocks \cite{b21,b13,b15}. Secondly, HS-TasNet \cite{b9} introduces a large number of redundant parameters (42 M), setting limit on its potential application scenarios. Thirdly, although HS-TasNet \cite{b9} is the state-of-the-art model in real-time, low-latency MSS, there is still considerable room for improvement, given the significant gap in terms of MUSDB18-HQ scores between it and the best models for offline MSS.

\begin{figure*}[t] 
\centering
  \includegraphics[width=\textwidth]{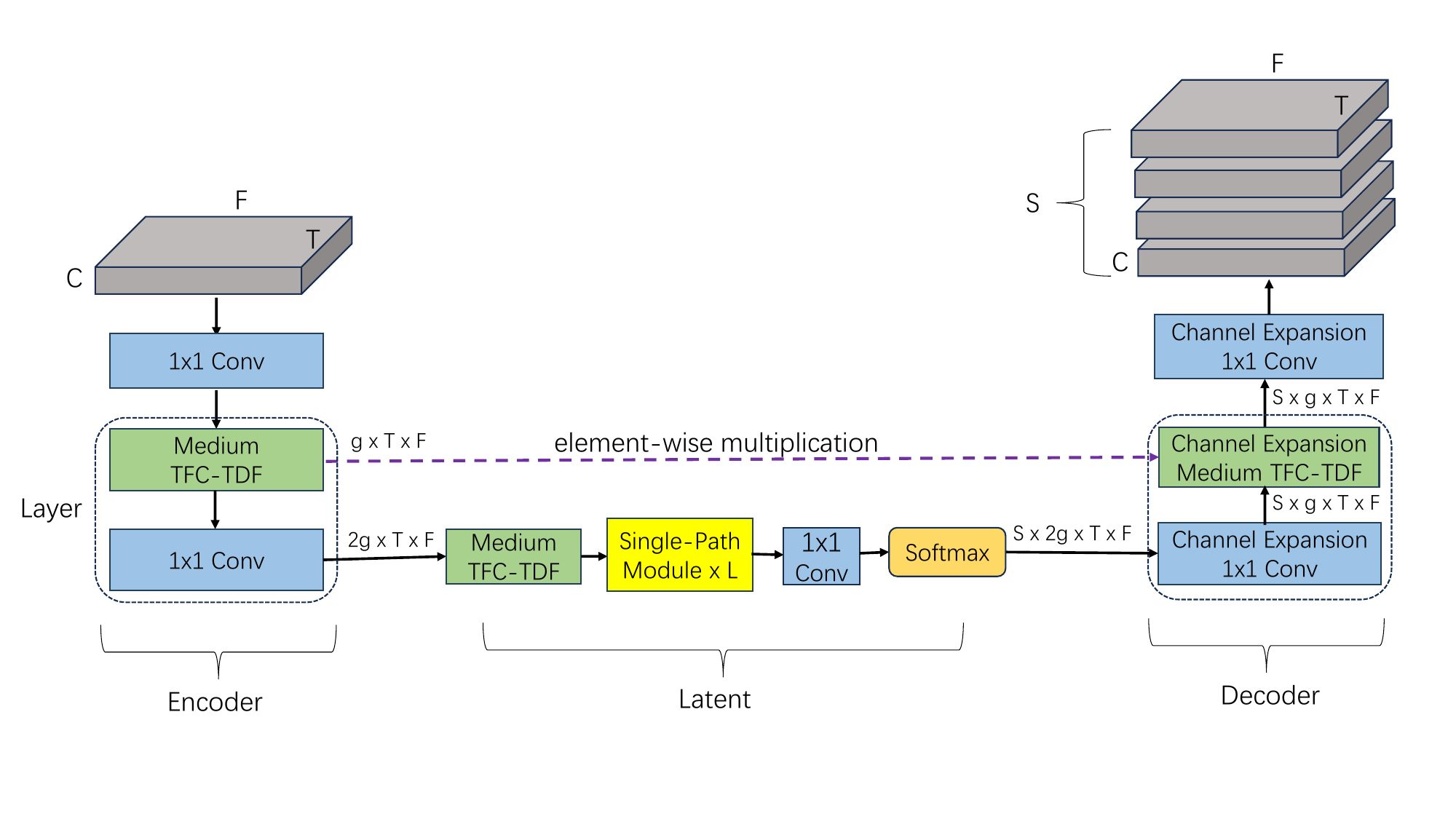}
  \caption{A framework of Real-Time Single-Path TFC-TDF UNet when the number of layers is one. F is the number of features on the frequency axis; T is the number of features on the time axis; C is the number of channels of input spectrogram; g indicates the channel increment; L is the number of Single-Path modules; S is the number of target sources.}
  \label{fig:a}
\end{figure*}

\begin{figure}
    \centering
    \includegraphics[width=0.5\linewidth]{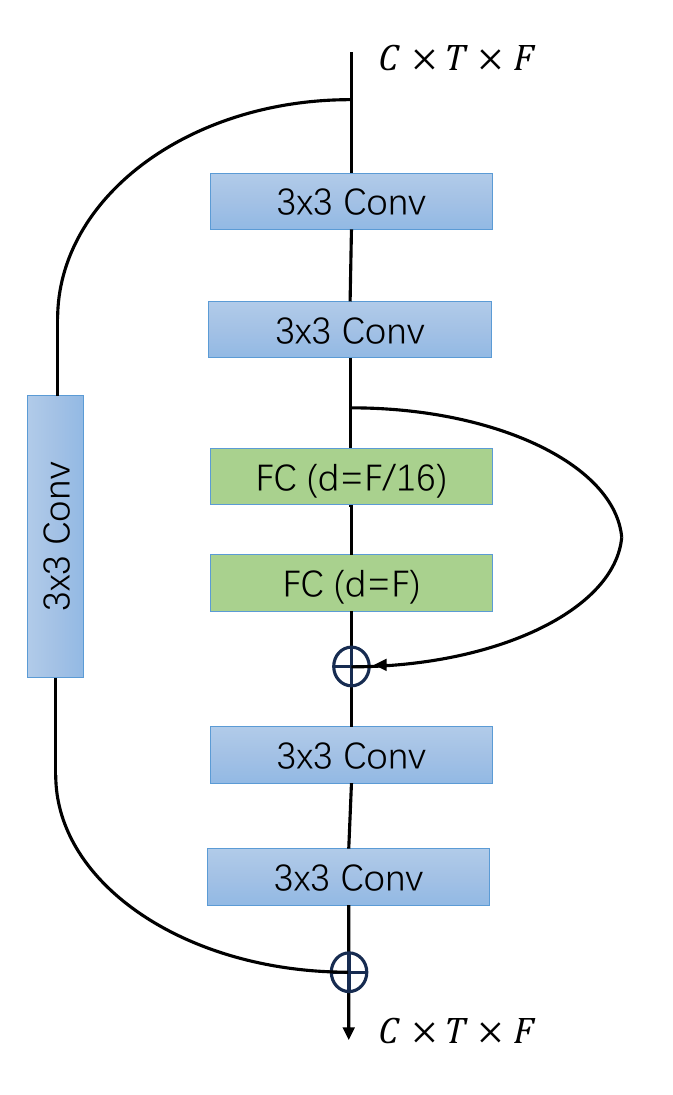}
    \caption{Medium TFC-TDF. F is the number of features on the frequency axis; T is the number of features on the time axis; C is the number of channels generated by the convolution layer; d is the size of the output features for the fully connected layer.}
    \label{fig:b}
\end{figure}

\section{Methodolgy}
\subsection{RT-STT}

In this section, we present our RT-STT, as demonstrated in Fig. \ref{fig:a}. The RT-STT model is characterized by a shallow network. Element-wise multiplication is applied between the encoder and decoder, serving as skip connections, as indicated by the dotted arrows. Notably, all modules in RT-STT adopt a causal implementation in the time dimension.

\subsubsection{Encoder}
Let $\mathbf{x} \in \mathbb{R}^{C_0 \times L}$ denote the audio waveform of the input mixture, where $C_0$ and $L$ represent the number of channels and the number of audio samples, respectively. The module takes as input the real-valued spectrogram $\mathbf{X} \in \mathbb{R}^{C \times F \times T}$ generated by the short-time Fourier transform (STFT), with $C=2 \times C_0$, because the real and imaginary parts are separated into distinct channels. $F$ and $T$ denote the frequency and temporal dimensions, respectively. The encoder begins by applying a 1\texttimes1 convolution to increase the input channel from $C$ to $g$. Subsequently,  the architecture includes a single layer with a medium TFC-TDF block, followed by another 1\texttimes1 convolution that increases the channel count from $g$ to $2g$.

\subsubsection{Medium TFC-TDF block}
The medium TFC-TDF block, as depicted in Fig. \ref{fig:b}, is a streamlined version of the original TFC-TDF v3 block \cite{b21}, featuring fewer layers and a reduced latent feature size between the two fully connected layers. The input is first processed by a TFC block, which consists of two convolutional layers. Subsequently, a residual TDF block reduces and then recovers the size of the frequency dimension. This is followed by another TFC block. Finally, a residual connection from a single convolutional layer is added to the output.

\subsubsection{Latent Layer}
As shown in Fig. \ref{fig:a}, the latent layer first employs a medium TFC-TDF block, after which the single-path module is repeated $L$ times. The single-path module is further discussed in section \ref{sec:single-path}. Subsequently, a 1\texttimes1 convolution is employed to scale the input size according to the number of target sources, generating the source dimension. To enhance performance, the Softmax activation function is applied to the source dimension \cite{b11}.

\subsubsection{Decoder}
The decoder mirrors the encoder’s architecture but incorporates feature fusion based on channel-expansion convolution, which is detailed in section \ref{sec:feature fusion}.

\subsection{Feature Fusion Based on Channel Expansion}
\label{sec:feature fusion}
In this section, we discuss the method of feature fusion based on channel expansion in RT-STT. This method is straightforward to implement and can improve the performance of our model while reducing inference time.

Initially, in the decoder block, we perform independent convolutions for different sources. This type of convolution fully exploits the independence of each source, avoiding mutual interference and allowing for more dedicated modeling. However, the success of combining distinct instrument estimates in \cite{b10} provides valuable insights. In \cite{b10}, the original target network is built from sub-networks for each source, and it is demonstrated that bridging the network paths via a simple averaging operation enables each extraction network to better understand the others, thereby improving overall performance. Gaining insights from \cite{b10}, we utilize channel expansion convolution for feature fusion. Instead of performing separated convolutions for distinct sources, we combine the channels of different sources to facilitate a mixed convolution operation. In this way, the convolution layers are able to  capture interactions between different sources and jointly model distinct features. Additionally, employing feature fusion based on channel expansion reduces inference time, as modern hardware (GPU) parallelizes operations more efficiently across the channel dimension \cite{b37}.

As shown in TABLE \ref{tab:joint modeling}, the performance of the models is significantly enhanced through the utilisation of joint modeling, which implies that the mutual dependency between different sources outweighs the full exploitation of independence of individual sources in the field of MSS. The inference time also decreases by 0.2 ms, making it more suitable for low-latency applications.

\subsection{Single-Path Module}
\label{sec:single-path}
The single-path module comprises two RNN blocks. Each RNN block includes a group normalization layer, a Long Short-Term Memory (LSTM) unit, and a fully connected layer that restores the input shape. Finally, a residual connection is added to the output. This single-path module is derived from the dual-path module in DTTNet \cite{b13}. Unlike the dual-path module, which alternates between modeling the time and frequency dimensions through dimension reordering between two RNN blocks, the single-path module in RT-STT consistently focuses on modeling the time dimension.

In DTTNet, the dual-path module plays a crucial role, processing the time and frequency dimensions alternately, thereby facilitating both temporal and frequency modeling. However, this mechanism may function differently in real-time MSS models. To achieve low latency, real-time models typically employ smaller window sizes and more compact network architectures, which can compromise the effectiveness of the dual-path module in three key aspects.

Firstly, as the window size decreases, the frequency resolution diminishes, resulting in larger frequency bins that cover broader frequency ranges. Consequently, frequency information that should ideally be captured in finer detail becomes more blurred. In this scenario, modeling the frequency dimension, as performed by the dual-path module, becomes challenging because the lower frequency resolution hampers the ability to capture local structures and fine details in the frequency domain. As a result, the dual-path module’s performance in modeling frequency-related information may deteriorate. Secondly, as the window size decreases, the time resolution improves, allowing the STFT to capture more detailed temporal information in each frame. Under these circumstances, the changes in the temporal dimension become more pronounced, making temporal information more crucial. Therefore, a single-path module that focuses solely on temporal modeling may outperform the dual-path module because it can better capture dynamic changes in the time domain without needing to account for the relatively less relevant or less clear frequency dimension. Thirdly, a reduction in network depth implies a decrease in overall model complexity. The dual-path module, which alternates between modeling time and frequency, inherently adds complexity to the network. However, as the network depth decreases, this alternating modeling approach may no longer perform as effectively as simpler, temporal-only modeling approaches. With reduced network capacity, the dual-path module may struggle to capture the intricate interactions between time and frequency, whereas the single-path module, which exclusively models the temporal dimension, is relatively simpler and more effective. 

In addition to the above three reasons related to model performance, dual-path module is also more time-intensive. Due to the additional time needed for reordering dimensions when modeling frequency and time alternately, the inference time of the dual-path module is considerably higher than that of the single-path module.

To compare the effectiveness of dual-path and single-path modules in real-time MSS models, we conduct an experiment. As shown in TABLE \ref{tab:single-path}, the scores improve slightly when we consistently perform time modeling, with a reduced inference time. This verifies the superiority of single-path modeling over dual-path modeling in real-time applications.

\subsection{Quantization}

\begin{table*}[t]
\centering
\caption{Comparing RT-STT with state-of-the-art real-time models and some famous offline models. All models listed in this table are trained only on MUSDB18-HQ without extra data except the non-causal Tasnet$^{+}$ (using 2,500 extra songs). The original X-UMX \cite{b10}, DemucsV2 \cite{b20}, DemucsV4 \cite{b8} and DTTNet \cite{b13} cannot process inputs as small as 1,024 samples. Latency in this table represents algorithmic latency. For real-time models, the algorithmic latency is equivalent to the window size.}
\begin{tabular}{|c|c|c|c|c|c|c|c|c|c|}
\hline
 & \textbf{Architecture} & \textbf{All} & \textbf{Vocals} & \textbf{Drums} & \textbf{Bass} & \textbf{Other}  & \textbf{Latency (ms)} & \textbf{Param.}  & \textbf{Time on GPU (ms)} \\
\hline
 & X-UMX \cite{b10} & 5.79 & 6.61 & 6.47 & 5.43 & 4.64 & 8,000 & 36 M & - \\
 & TasNet$^{+}$ \cite{b19} & 6.52 & 7.34 & 7.68 & 7.04 & 4.04 & 15,000 & 29 M & \textbf{2.38} \\
\textbf{Non-causal} & DemucsV2 \cite{b20} & 6.28 & 6.84 & 6.86 & 7.01 & 4.42 & 8,000 & 288 M & - \\
\textbf{models} & DemucsV4 \cite{b8} & 7.52 & 7.93 & \textbf{7.94} & \textbf{8.48} & 5.72 & 7,800 & 41 M & - \\
 & DTTNet \cite{b13} & \textbf{8.06} & \textbf{10.12} & 7.74 & 7.45 & \textbf{6.92} & 6,000 & \textbf{20 M} & - \\
\hline
 & X-UMX \cite{b9} & 3.93 & 4.65 & 4.36 & 3.79 & 2.92 & 23 & 31 M & 1.80 \\
 & TasNet \cite{b9} & 4.40 & 5.02 & 4.38 & 4.73 & 3.48 & 23 & 51 M & 1.90 \\
\textbf{Real-time Low-} & HS-TasNet-Small \cite{b9} & 4.48 & 5.21 & 4.89 & 4.42 & 3.42 & 23 & 16 M & 2.10 \\
\textbf{latency models} & HS-TasNet \cite{b9} & 4.65 & 5.13 & 5.22 & 4.59 & 3.64 & 23 & 42 M & 3.90 \\
 & RT-STT (before quantization) & \textbf{5.17} & \textbf{5.56} & \textbf{5.84} & \textbf{5.26} & \textbf{4.02} & 23 & \textbf{383 K} & 5.80 \\
 & RT-STT (after quantization) & \textbf{5.17} & \textbf{5.56} & 5.83 & 5.25 & \textbf{4.02} & 23 & \textbf{383 K} & \textbf{1.01} \\
\hline
\end{tabular}
\label{tab2}
\end{table*}

\begin{table}[t]
\caption{cSDR in dB on MUSDB18HQ and inference time on 1024 samples (2 channels) for RT-STT with dual-path modules or single-path modules.}
\begin{center}
\begin{tabular}{|c|c|c|c|c|c|c|}
\hline
$path$ & Vocals & Drums & Bass & Other & All &  $T_p (ms)$ \\
\hline
dual & \textbf{5.55} & 5.78 & 5.23 & \textbf{4.10} & 5.16 & 6.7 \\
\hline
single & \textbf{5.55} & \textbf{5.84} & \textbf{5.26} & 4.02 & \textbf{5.17} & \textbf{5.8} \\
\hline
\end{tabular}
\label{tab:single-path}
\end{center}
\end{table}

\begin{table}[t]
\caption{cSDR in dB on MUSDB18HQ and inference time on 1024 samples for RT-STT with and without joint modeling. $JM$ denotes whether the model employs joint modeling. $T_p$ indicates the inference time on input of 1024 samples (2 channels) averaged across 1000 iterations.}
\begin{center}
\begin{tabular}{|c|c|c|c|c|c|c|}
\hline
$JM$ & Vocals & Drums & Bass & Other & All &  $T_p (ms)$ \\
\hline
$\times$ & 5.21 & 4.91 & \textbf{5.66} & 3.87 & 4.92 & 6.0 \\
\hline
\checkmark & \textbf{5.55} & \textbf{5.84} & 5.26 & \textbf{4.02} & \textbf{5.17} & \textbf{5.8} \\
\hline
\end{tabular}
\label{tab:joint modeling}
\end{center}
\end{table}

\begin{table}[t]
\caption{Hyper-parameters and Performance. All the models are trained on MUSDB18-HQ without extra data. All the models adopt window size of 1024 samples and hop size of 512 samples. 'layers' indicates the depth of network; 'g' indicates the channel increment; 'L’ indicates the repeated times of single-path module; cSDR is the average cSDR for 4 tracks. Inference time is calculated on input of 1024 samples averaged across 1000 iterations.}
\begin{center}
\begin{tabular}{|c|c|c|c|c|c|c|c|c|c|}
\hline
\textbf{layers } & \textbf{ g } & \textbf{ L } & \textbf{ cSDR } & \textbf{ Inference time } \\
\hline
1 & 16 & 3 & 5.17 &  5.80 ms \\ 
\hline
2 & 16 & 3 & 5.51 &  12.11 ms \\ 
\hline
2 & 8 & 3 & 5.16 &  9.85 ms \\
\hline
1 & 16 & 4 & 5.18 &  6.02 ms \\
\hline
\end{tabular}
\label{tab:last}
\end{center}
\end{table}

Quantization of a deep neural network refers to the process of reducing the precision of the numerical values used in the model’s parameters and the intermediate activations during computation \cite{b22}. The primary goal of quantization is to reduce the memory footprint and computational requirements of the model without significantly sacrificing performance. Although quantization is widely used in computer vision \cite{b23,b24}, there has been a paucity of literature examining its impact on audio separation. In our experiment, we conduct Post-Training Quantization (PTQ) as well as Quantization-Aware Training (QAT) , achieving an 82.6\% reduction in inference time without a significant decrease in performance. These results provide useful insights such as quantization not compromising performance but rather enhancing the inference speed.

We first apply PTQ to our model, aiming to reduce the precision of RTT to 16-bit floating point. The PTQ tools we use are Open Neural Network Exchange (ONNX), a format designed to facilitate the interoperability of deep learning models across different frameworks, and TensorRT, a high-performance deep learning inference library designed to accelerate the inference of deep learning models.
Since we have consistently been training our model using mixed-precision training, the PTQ process proceeds smoothly. As demonstrated in the last two lines in TABLE \ref{tab2}, we reduce the inference time from 5.8 ms to 1.01 ms without a significant drop in performance (within 0.01). 

When we combine QAT and PTQ to reduce the precision to 8-bit integer, however, the inference time only decreases to 1.06 ms. It can be surmised that this phenomenon is associated with the provision of hardware support, the patterns of memory access, and the compensation for quantization errors. Furthermore, the quantization to 8-bit integer results in a more significant performance degradation, so ultimately we adopt the quantization to 16-bit floating point.

\section{Experiment}
\subsection{Dataset}
The MUSDB18-HQ dataset \cite{b12} comprises 150 songs, each sampled at 44,100 Hz with 2 channels. Each song contains 4 independent tracks: ’vocals’, ’drums’, ’bass’, and ’other’. The training set consists of 86 songs, the validation set comprises 14 songs, and the testing set includes 50 songs. For data augmentation, we apply pitch shifts of \{-2, -1, 0, 1, 2\} semitones and time stretches of \{-20\%, -10\%, 0\%, 10\%, 20\%\}.

\subsection{Experimental Setup}
Our training pipeline is based on that of DTTNet \cite{b13}, with the following modifications: the learning rate is set to $1 \times 10^{-4}$, gradient clipping is utilized with a maximum L2 norm of 3 for the gradient, and the chunk size is set to 32,256. Additionally, we use four RTX 3090 GPUs, each with a batch size of 8. For fine-tuning, we add L1 normalization.

For the STFT, we use a window size of 1,024 and a hop length of 512. Inspired by \cite{b15}, we further cut the frequency bins to 384. Therefore, during training, we have $F = 384$ and $T = 64$. For the Single-Path Module, we set $L = 3$. The channel increment factor $g$ is set to 16 and $S$ is set to 4 as shown in Fig. \ref{fig:a}.

\subsection{Evaluation}
The SDR is employed for performance measurement.
\subsubsection{Chunk-level SDR (cSDR)} cSDR calculates the SDR on 1-second chunks and reports the median value. This metric was served as the default evaluation metric in the Signal Separation Evaluation Campaign (SiSEC) \cite{b16}.
\subsubsection{Utterance-level SDR (uSDR)} uSDR calculates the SDR for each song and reports the mean value. This metric was used as the default evaluation metric in the Music Demixing (MDX) Challenge 2021 \cite{b2}. The definition of uSDR is identical to the standard stereo signal-to-noise ratio (SNR).

\section{Results}

\subsection{Hyper-parameters and Performance}
In this section, we experiment with different combinations of hyper-parameters to study their impact on cSDR and inference time, aiming to achieve an optimal balance between separation quality and inference speed. The results are presented in TABLE \ref{tab:last}.

\subsubsection{Number of Layers and Channel Increment Factor}
As shown in the first and second rows of TABLE \ref{tab:last}, increasing the number of layers improves separation quality from 5.17 dB to 5.51 dB, but the additional inference time from 5.80 ms to 12.11 ms is unacceptable. Although we can decrease the channel increment factor $g$ to accelerate inference, as shown in the third row, both the separation quality and inference speed are inferior to those of the model with the hyper-parameters listed in the first row.

\subsubsection{Repeated Times of Single-Path Module}
As shown in the first and last rows of TABLE \ref{tab:last}, when we increase the number of repeated times of the single-path module from 3 to 4, cSDR only experiences a minor increase from 5.17 dB to 5.18 dB, with inference time increasing by 0.22 ms. Consequently, setting $L = 3$ is sufficient to achieve good separation quality and inference time as compared to $L = 4$.

\subsection{Comparison with the State-of-the-art}
The results are shown in TABLE \ref{tab2}. To compare the inference speed of our model with the baseline \cite{b9}, we present the inference time of the models to process 1024 samples (two channels) on NVIDIA RTX 3080Ti GPU. For non-causal models, the original X-UMX \cite{b10}, DemucsV2 \cite{b20}, DemucsV4 \cite{b8} and DTTNet \cite{b13} cannot process an input as small as 1024 samples, so TABLE \ref{tab2} doesn't contain their inference time. Notably, the non-causal TasNet is unable to run in real time, but we still report the time required to process 1024 samples for illustrative purposes \cite{b9}. For real-time low-latency models, we can see that, compared with real-time TasNet \cite{b9}, HS-TasNet increases average cSDR by 0.25 dB at the cost of an additional 2 ms. In contrast, our RT-STT significantly enhances the average cSDR from 4.65 dB (HS-TasNet) to 5.17 dB, with inference time increasing by 1.9 ms before quantization. Moreover, quantization has enabled a reduction in processing time on the GPU to a minimum of 1.01 ms. Regarding parameters, our RT-STT reduces the number of parameters from 20 M (DTTNet) to 383 K. Compared with other real-time models, our model minimizes redundant parameters, thereby paving the way for potential applications.

\section{Conclusion}
Although significant progress has been made in the field of deep learning for music demixing, limited attention has been paid to real-time, low-latency music demixing. In this work, we introduce a lightweight real-time low-latency model called RT-STT, which outperforms the state-of-the-art model in multiple aspects, including MSS performance, inference speed, and model size. Apart from that, we implement feature fusion based on channel expansion and demonstrate the clear advantages of single-path modeling over dual-path modeling in real-time applications. Furthermore, we apply quantization techniques to accelerate the inference process, showing that this approach maintains performance while significantly reducing computational cost. In future work, we aim to enhance the model’s ability to handle higher audio quality and further reduce computational cost, making it suitable for deployment on some low-resource, low-power platforms.


\bibliographystyle{IEEEbib}
\bibliography{icme2025references}

\end{document}